\documentclass[aps,prl,twocolumn,superscriptaddress]{revtex4}

\usepackage{amsmath}
\usepackage{amssymb}
\usepackage{amsfonts}
\usepackage{mathrsfs}
\usepackage{graphicx}

\begin{document}

\title{Manipulating single enzymes by an external harmonic force}

\author{Michael A. Lomholt}
\affiliation{Physics Dept, University of Ottawa, 150 Louis Pasteur, Ottawa,
ON, K1N 6N5, Canada}
\author{Michael Urbakh}
\affiliation{School of Chemistry, Tel Aviv University, 69978 Tel Aviv,
Israel}
\author{Ralf Metzler}
\affiliation{Physics Dept, University of Ottawa, 150 Louis Pasteur, Ottawa,
ON, K1N 6N5, Canada}
\affiliation{NORDITA, Blegdamsvej 17, DK-2100 Copenhagen {\O}, Denmark}
\author{Joseph Klafter}
\affiliation{School of Chemistry, Tel Aviv University, 69978 Tel Aviv,
Israel}

\begin{abstract}
We study a Michaelis-Menten reaction for a single two-state enzyme molecule,
whose transition rates between the two conformations are modulated by an
harmonically oscillating external force. In particular, we obtain a range of
optimal driving frequencies for changing the conformation of the enzyme thereby
controlling the enzymatic activity (i.e. product formation). This analysis
demonstrates that it is, in principle, possible to obtain
information about particular rates within the kinetic scheme.
\end{abstract}

\maketitle

Recent advances in single molecule spectroscopy have made it possible
to follow the catalytic activity of individual enzymes over extended
periods of time \cite{edman00,schenter99,english06,kou05,flomenbom05,lerch05}. Although
the catalytic rates have been shown to be distributed in time, the
general Michaelis-Menten scheme, originally proposed for an ensemble
of enzymes, turned out to hold quite well also on the level of a single enzyme
\cite{edman00,schenter99,english06,kou05,flomenbom05,lerch05,min06,xue06}. What is still left
open is the nature of the conformational changes and its relationship to
the enzymatic activity \cite{lerch05,min06,xue06,flomenbom06}. Deeper insights
into the conformational-activity
relationship can be obtained by influencing single enzyme conformations by
an external force and following the effects on the catalytic activity. The
possibility to manipulate enzymatic turnovers by applying a mechanical
force has been recently demonstrated in \cite{wiita06}. Also, voltage-gated ion
channels can be manipulated electrically to switch between two states \cite{pustovoit06}. Here, assuming a two-state Michaelis-Menten scheme for a single enzyme, we
investigate how one can manipulate the activity (turnovers) of an enzyme
by an external harmonic force. The oscillatory force is shown to be able to
increase the weight (occurrence time) of a desired conformation in the
scheme, controlling the enzyme activity.

\begin{figure}[t]
\includegraphics{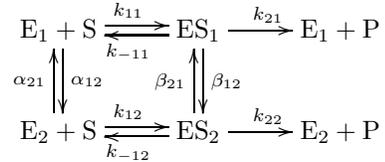}
\caption{Michaelis-Menten scheme for a single enzyme with two conformations.
E denotes the enzyme molecule, S the substrate, and P the product. Rates $\alpha_{ij}$ and
$\beta_{ij}$ ($i,j=1,2$) stand for transitions between conformational states 1 and 2, 
while rates $k_{xi}$ ($x=-1,1,2$) quantify internal conversions.}
\label{fig:schemefig}
\end{figure}

The reaction scheme that we examine is shown in
Fig.~\ref{fig:schemefig}.
We model the effect of the time-dependent external force as a shift in the
conformational energy landscape. Assuming that the transitions between
different conformations, with rates $\alpha_{ij}$ and $\beta_{ij}$, follow
an Arrhenius activation, and using a harmonic
force which modulates barriers separating conformational states,
we find the following time dependence of the rates
\begin{subequations}
\begin{eqnarray}
\alpha_{ij}&=&\alpha_{ij}(t)=\alpha_{ij}^{(0)}\exp\left(-\epsilon_{ij}^{
(\alpha)}\sin(\omega t+\phi_0)\right),\\
\beta_{ij}&=&\beta_{ij}(t)=\beta_{ij}^{(0)}\exp\left(-\epsilon_{ij}^{(\beta)}
\sin(\omega t+\phi_0)\right).
\end{eqnarray}\label{eq:eqs1}
\end{subequations}
Here, $\omega$ is the angular frequency of the external driving force,
$\phi_0$ the initial
phase of the force; and $\alpha_{ij}^{(0)}$, $\beta_{ij}^{(0)}$, $\epsilon
_{ij}^{(\alpha)}$, and $\epsilon_{ij}^{(\beta)}$ are phenomenological
constants characterizing the enzyme and the amplitude of the driving force.
In Eqs. (\ref{eq:eqs1}) we assume that the potential energy landscape of the
conformations responds instantly to changes in the force. This assumption can be
relaxed, for instance by including different phases $\phi_0$ for $\alpha_{ij}$
and $\beta_{ij}$. Note, however, that a $180^\circ$ phase difference,
corresponding to a change of sign for $\epsilon_{ij}^{(\beta)}$, is included
in Eqs. (\ref{eq:eqs1}). We assume that
the substrate S is in great excess, such that the rates for the reaction with the substrate
like $k_{1i}=k_{1i}^{(0)}
[{\rm S}]$ become independent of time. The scheme in Fig. \ref{fig:schemefig}
is closely related to schemes describing resonant activation in discrete systems
\cite{flomenbom04}.

\begin{figure}
\includegraphics[width=7.2cm]{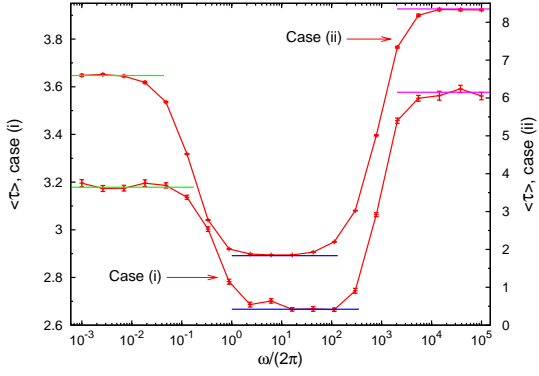}
\caption{Mean turnover time $\left<\tau\right>$ as function of driving
frequency $\omega$ for the case $k_{xi}\ll\alpha_{ij},\beta_{ij}$. Horizontal
lines represent the limit cases derived in the text. (i) $k_{1i}=
k_{-1i}=1$, $k_{21}=0.3$, $k_{22}=3$, $\alpha_{12}^{(0)}=80$,
$\alpha_{21}^{(0)}=800$, $\beta_{12}^{(0)}=160$, $\beta_{21}^{(0)}=1600$,
$\epsilon_{12}^{(\alpha)}=2=-\epsilon_{21}^{(\beta)}$, $\epsilon_{12}^{
(\beta)}=3=-\epsilon_{21}^{(\alpha)}$.
(ii) the following parameters were changed: $k_{11}=k_{-11}=k_{-12}=0.01$,
$k_{12}=3=-\epsilon_{12}^{(\beta)}$, $\epsilon_{21}^{(\beta)}=2$ (units on right axis).
Simulations were run for $5\times 10^4$ turnovers per frequency.}
\label{fig:fasttav}
\end{figure}

\begin{figure}[tb]
\includegraphics[width=7.2cm]{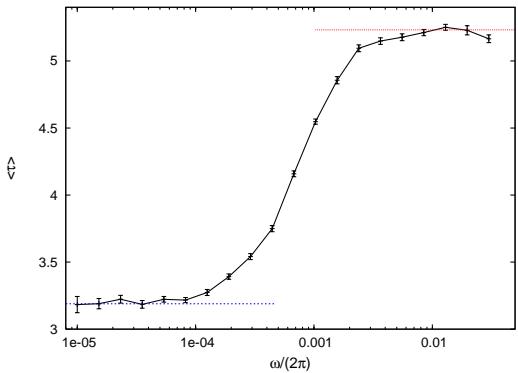}
\caption{Plot of $\left<\tau\right>$ as function of driving frequency $\omega$
for $\alpha_{ij},\beta_{ij}\ll k_{xi}$. The horizontal lines represent the
limits derived
in the text. The parameters are $k_{1i}=k_{-1i}=1$, $k_{21}=0.3$,
$k_{22}=3$, $\alpha_{12}^{(0)}=2\times 10^{-4}$, $\alpha_{21}^{(0)}=
10^{-3}$, $\beta_{12}^{(0)}=2\times 10^{-5}$, $\beta_{21}^{(0)}=
10^{-4}$, $\epsilon_{12}^{(\alpha)}=3$, $\epsilon_{21}^{(\alpha)}=-3$,
$\epsilon_{12}^{(\beta)}=3$ and $\epsilon_{21}^{(\beta)}=-3$. The simulation
comprises $2\times 10^6$ turnovers per frequency.}
\label{fig:slowtav}
\end{figure}

\emph{Simulations.} As shown below, for certain limits of the rates analytical
expressions can be derived. To obtain insight into the general behavior
of enzymatic activity including noise, we use a stochastic simulation. To
this end, we implemented the Gillespie algorithm
\cite{gillespie}, that provides random
values for the waiting time between two reaction steps, and the direction
of the reaction as weighted by the corresponding Arrhenius factor. We start the system in one of the states ${\rm E}_i+{\rm
S}$, and then perform jumps to one of the neighboring states according to the
Gillespie reaction probability. Which state it jumps to is decided by picking a
waiting time for jumping to state ${\rm ES}_i$ according to the exponential
distribution with rate parameter $k_{1i}$, and one for jumping to ${\rm E}_{j}
+{\rm S}$ according to the cumulative distribution
\begin{equation}
F_{\alpha_{ij}}(\Delta t)=1-\exp\left(-\int_t^{t+\Delta t} \alpha_{ij}(t')d t'
\right),
\label{eq:alphacumu}
\end{equation}
and then the jump is made to the state with the shortest waiting
time. Subsequent jumps to neighboring states are determined similarly, and one
iteration of the simulation stops at time $t=\tau$ when the system reaches one
of the product states ${\rm E}_i+{\rm P}$. The conformation $i$ of the enzyme
is then used as the initial state for the next iteration with the
initial phase $\phi_0^{({\rm new})}=\omega\tau+\phi_0^{({\rm old})}$ determining the new
momentary force. Performing iterations over many periods of the oscillating force we obtain a
time series for the turnover times $\tau$, from which the average waiting time
$\left<\tau\right>$ yields. Plots of $\left<\tau\right>$ as a function of
$\omega$ are shown in Figs.~\ref{fig:fasttav} and \ref{fig:slowtav}.
Depending on the choice of parameters, switching between two or three kinetic regimes of
the system is revealed. Certain limiting behaviors are accessible analytically,
as discussed soon. For the parameters used
throughout this work the rate of product formation in the conformation 2 is higher than
in conformation 1.

We note that in our analysis the parameters (except for case (ii) in
Fig. \ref{fig:fasttav}) are chosen such that
detailed balance is fulfilled, as required for certain single enzymes
that are coupled to the surrounding heat bath \cite{gopich06}. The explicit
condition arising when detailed balance is applied around the loop in
the reversible reaction pathway in Fig.~\ref{fig:schemefig}, is
$k_{-11}\beta_{21}(t)k_{12}\alpha_{12}(t)=k_{11}\beta_{12}(t)
k_{-12}\alpha_{21}(t)$.
The condition is applied for any momentary value of the external force, and
therefore any time $t$, since the system could equally well be held constantly
at these forces.
Detailed balance could be violated, for instance, by enzymes converting ATP energy during
their cycle.

\begin{figure}
\includegraphics[width=7.2cm]{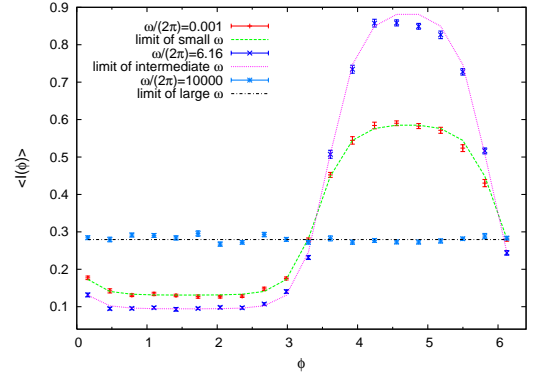}
\caption{Plot of $\left<I(\phi)\right>$ as a function of $\phi$ for different
driving frequencies $\omega$, for the same conditions as in
Fig.~\ref{fig:fasttav} (i).}
\label{fig:fastphase}
\end{figure}

\emph{Fast switching limit.} Consider the case when the switching between
enzyme conformations is much faster than the reactions with the substrate,
as displayed in Fig.~\ref{fig:fasttav}. Note that case (ii) in Fig.~\ref{fig:fasttav} has different $k_{1i}$'s \cite{thefootnote}. It
also includes a $180^\circ$ shift in the phases of 
variation of rates $\alpha_{ij}$ and $\beta_{ij}$ under
the external force distinguishing between substrate-bound and substrate-free
enzyme states. Parameters were chosen to reflect situations with enhanced
enzyme efficiency at intermediate frequencies. Different scenarios, for
instance with decreased efficiency, are also possible. In both cases in
Fig. \ref{fig:fasttav} the
rates satisfy $k_{xi}\ll\alpha_{ij},\beta_{ij}$. This means that we can define
effective rates $k_x$ as averages over the various conformations. The weights
in these averages are given by the probabilities for the enzyme to be in the
different conformations. If $\omega$ is much faster or slower than the
relaxation time between the two enzyme conformations, these
probabilities can be easily obtained. For instance, the probability for the
enzyme to be in conformation ${\rm E}_1$, given that no substrate is bound, is
\begin{equation}
p(1|E,\phi)=\left\{\begin{array}{ll}
\alpha_{21}(\phi)/[\alpha_{12}(\phi)+\alpha_{21}(\phi)], &\omega\ll
\alpha_{ij}\\[0.2cm]
{\bar \alpha}_{21}/[{\bar \alpha}_{12}+{\bar \alpha}_{21}], &\omega\gg
\alpha_{ij}
\end{array}\right.
\label{eq:piEphi}
\end{equation}
where $\phi\in(0,2\pi)$ is the phase of the force, i.e., $\omega t+\phi_0=
\phi+2\pi n$ for some integer $n$, and a bar over a quantity means that this
quantity is averaged over all phases,
\begin{equation}
{\bar \alpha}_{ij}=\int_0^{2\pi}\frac{d \phi}{2\pi}\alpha_{ij}(\phi)=
\alpha_{ij}^{(0)}I_0\left(\epsilon_{ij}^{(\alpha)}\right),
\end{equation}
where $I_0$ is the modified Bessel function of order 0.
The effective rates $k_x$ can now be obtained in the two limits,
\begin{equation}
k_x(\phi)=\left\{\begin{array}{ll}
\sum_i k_{xi}p(i|x,\phi),&\omega\ll k_{xi}\\[0.2cm]
\sum_i k_{xi}{\bar p}(i|x),&\omega\gg k_{xi}
\end{array}\right.,\label{eq:kxphi}
\end{equation}
where $x$ in $p(i|x,\phi)$ refers to the unbound state ${\rm E}+{\rm S}$ for $x=1$, and to the bound state ${\rm ES}$ for $x=-1$ and $x=2$.
The rate of the overall reaction is then given by the standard
Michaelis-Menten expression \cite{kou05,english06,gopich06}
\begin{equation}
\nu(\phi)=\frac{k_2(\phi)k_1(\phi)}{k_1(\phi)+k_{-1}(\phi)+k_2(\phi)}
\label{eq:nuphifast}
\end{equation}
as function of the phase $\phi$. The average waiting time $\left<\tau\right>$
for one turnover can now be found by averaging over all phases and calculate
$\left<\tau\right>=1/{\bar \nu}$. The values of $\left<\tau\right>$ calculated
according to Eqs.~(\ref{eq:kxphi}) and (\ref{eq:nuphifast}) are shown in
Fig.~\ref{fig:fasttav} by the horizontal lines. We note that the limits where
$\omega\ll \alpha_{ij},\beta_{ij}$ can also be obtained as limit cases of
the system studied in Ref.~\cite{astumian93}.

Another experimentally relevant quantity is the temporal probability density
for the formation of products at a given value of the force as determined by
the phase $\phi$. A plot of this intensity of product formation events, which
we label by $\left<I(\phi)\right>$, is shown in Fig.~\ref{fig:fastphase}. An approximate
bimodal behavior is found for certain choices of the parameters. To obtain
the analytic limits we note that if the external force is slowly varying,
i.e., $\omega\ll k_{xi},\alpha_{ij},\beta_{ij}$, then the intensity will
simply be equal to the rate $\left<I(\phi)\right>=\nu(\phi)$. This
expression also holds for $\omega\gg \alpha_{ij},\beta_{ij}$
where it is independent of $\phi$. For the intermediate $\omega$
where $k_{xi}\ll \omega \ll \alpha_{ij},\beta_{ij}$ it will be the
distribution of conformations in the final step of the reaction and
the corresponding rate constants that determines when the product
formation happens, such that $\left<I(\phi)\right>={\bar \nu}\sum_i
k_{2i}p(i|{\rm ES},\phi)/{\bar k}_2$. We point out here 
that at slow frequencies it is the overall efficiency of the enzyme 
that is probed at different magnitudes of the external force, while 
at the intermediate frequencies it is only the relative efficiency 
of the final step in the reaction which is probed at different 
external driving. Thus tuning the frequency allows one to selectively 
study the final reaction step.

\begin{figure}[tb]
\includegraphics[width=7.2cm]{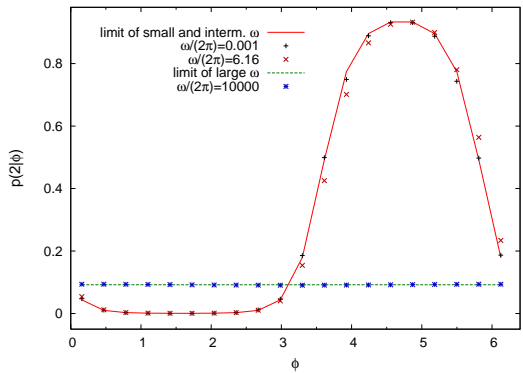}
\caption{Estimate of $p(2|\phi)$ as a function of $\phi$ for the same
parameters as in Fig.~\ref{fig:fasttav} (i). Note that there are only two
different analytic limits, since two of the three limits are identical for
the chosen parameters.}
\label{fig:fastprobphase}
\end{figure}

The behavior of $\left<I(\phi)\right>$ in Fig.~\ref{fig:fastphase} can be compared with
the probability of being in, say, conformation 2 (the fast one) as a function of $\phi$, as
shown in Fig.~\ref{fig:fastprobphase}. Note the large change in amplitude
for the chosen parameters. To obtain the analytic limits
we write the probability as $p(i|\phi)=p(i|{\rm E},\phi)p({\rm
E}|\phi)+p(i|{\rm ES},\phi)[1-p({\rm E}|\phi)]$, where $p(i|{\rm E},\phi)$
is given by Eq.~(\ref{eq:piEphi}). To find $p(E|\phi)$ note that at a given
phase $\phi$ the effective rate for the system to leave the state without
bound substrate is $k_{1}(\phi)$, and the rate for leaving the state with
bound substrate is $k_{-1}(\phi)+k_{2}(\phi)$. In fact the quasi-steady-state
probability for being in the state without bound substrate is
\begin{equation}
p(E|\phi)=\frac{k_{-1}(\phi)+k_{2}(\phi)}{k_1(\phi)+k_{-1}(\phi)+k_{2}(\phi)}
\label{eq:pEphi}
\end{equation}
in the two cases given by Eq. (\ref{eq:kxphi}) where either $\omega\ll k_{xi}$ or $\omega\gg k_{xi}$.
The resemblance of Figs. \ref{fig:fastphase} and \ref{fig:fastprobphase} reflects the dominating role of the conformation 2, due to its larger catalytic rate, in the enzymatic
activity.

\begin{figure}[tb]
\includegraphics[width=7.2cm]{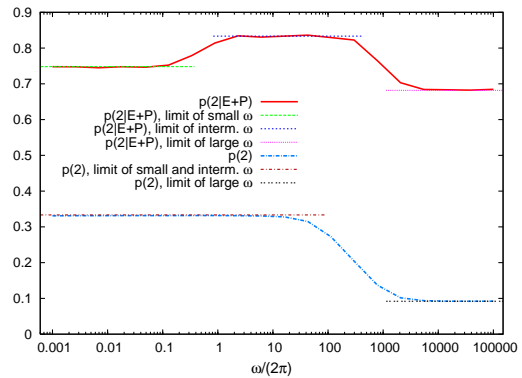}
\caption{Plot of the fraction $p(2)$ the enzyme spends in conformation 2 (lower curve),
as a function of driving frequency $\omega$ for the same parameters used
in Fig.~\ref{fig:fasttav} (i). The upper curve depicts the fraction
of product formation while the enzyme is in conformation 2.}
\label{fig:fastprob}
\end{figure}

To illustrate further the mechanisms behind the transitions of
$\left<\tau\right>$ with changing driving frequency we plotted the fraction of
time the enzyme spends in conformation 2, labeled $p(2)$, during a simulation
run in Fig.~\ref{fig:fastprob}.
The analytic limits of $p(2)$ can be obtained
straightforwardly by averaging the probability of being in conformation
2 over all phases, $p(i)=\int_0^{2\pi}[d\phi/(2\pi)]\;p(i|\phi)$. This
behavior can be compared with the fraction of product formation
events that occurs while the enzyme is in conformation 2, labeled
by $p(2|E+P)$ in Fig.~\ref{fig:fastprob}. The analytical limits
of $p(2|E+P)$ can be obtained by averaging over all phases $\phi$
the effective rate constant for forming a product in conformation $i$,
$k_{2i}p(i|{\rm ES},\phi)$, divided by the overall effective $k_2(\phi)$,
taking into account the varying rate of product formation by a factor
$\nu(\phi)/{\bar \nu}$. The explicit formula is
\begin{equation}
p(i|{\rm E+P})=\int_0^{2\pi}\frac{d\phi}{2\pi}\;\frac{k_{2i}p(i|{\rm
ES},\phi)}{k_2(\phi)}\,\frac{\nu(\phi)}{{\bar \nu}}.
\end{equation}
As Fig.~\ref{fig:fastprob} illustrates, the slowing down of the reaction with
increasing frequency at the second transition in Fig.~\ref{fig:fasttav}
occurs because the high frequency of the force shifts the enzyme towards
spending more time in the low efficiency conformation. The increased efficiency
of the enzyme at intermediate frequencies sets in when the external force
drives the enzyme back and forth between its two conformations fast enough
to avoid the bottleneck of the enzyme spending long uninterrupted periods
being forced towards the slowly reacting conformation.

\emph{Slow switching limit.} When the switching between enzyme states is slow
in comparison to substrate reactions, $\alpha_{ij},\beta_{ij}\ll k_{xi}$, the enzyme 
will mostly stay in the same
state during a reaction cycle, and we can therefore take the instantaneous
reaction rate to be
\begin{equation} \nu(\phi)=p(1|\phi)\nu_{1}+p(2|\phi)\nu_{2},
\end{equation}
where $\nu_{i}=k_{2i}k_{1i}/(k_{1i}+k_{-1i}+k_{2i})$ are the fixed
conformation rates and $p(i|\phi)$ is the probability for the enzyme
to be in conformation $i$ at a given phase of the oscillating force.
To find this probability we first argue in the same way
as Eq.~(\ref{eq:pEphi}) was obtained to see that the (steady-state) probability
of being in the state without bound substrate, given that the conformation is
$i$, is \begin{equation} p(E|i)=\frac{k_{-1i}+k_{2i}}{k_{1i}+k_{-1i}+k_{2i}}\;.
\end{equation} Then we note that the effective rate constant
for changing conformation from $i$ to $j$ can be found as
\begin{equation}
\gamma_{ij}=p(E|i)\alpha_{ij}+(1-p(E|i))\beta_{ij}.
\end{equation}
This effective rate now allows the
probability of being in conformation $i$ to be calculated similarly to
Eq.~(\ref{eq:piEphi}) as
\begin{equation}
p(i|\phi)=\left\{\begin{array}{ll}
\gamma_{ji}(\phi)/[\gamma_{ij}(\phi)+\gamma_{ji}(\phi)],&\omega\ll\gamma_{ij}\\ {\bar \gamma}_{ji}/[{\bar
\gamma}_{ij}+{\bar \gamma}_{ji}],&\omega\gg\gamma_{ij} \end{array}\right.
\end{equation}
where $j=3-i$. The averaged waiting time for one turnover
can again be found as $\left<\tau\right>=1/{\bar \nu}$.

\emph{Discussion.} Detailed knowledge of the dynamics of single enzymes
as well as their response to external stimulus is important for the
understanding of biochemical processes occurring in living cells. We
studied the response of a single two-state enzyme to a harmonic
external driving force in the presence of thermal noise. By combination
of analytic and simulations results, we demonstrated the rich response
behavior of the turnover dynamics of the enzyme to an external force. We
showed that in some cases there exists optimum driving frequencies that
minimize the turnover time of the enzyme. In these cases
one can selectively obtain
information on the reaction rates in the final step of the Michaelis-Menten
reaction by choosing an oscillation in a suitable frequency range. Moreover,
we found that the response of the enzyme activity can be very sensitive within
a small range of phase $\phi$, a signature of many biological switches. We
note that it should be possibly to access the full spectrum of relevant
driving frequencies by combining different experimental methods, such as
atomic force microscopy and optical switching methods.

MAL and RM acknowledge partial funding from the Natural Sciences and
Engineering Research Council (NSERC) of Canada, and the Canada Research
Chairs program. MU and JK acknowledge financial support from
the Deutsche Forschungsgemeinschaft (HA 1517/26-1,2 Single molecules).


\begin{thebibliography}{99}
\bibitem{thefootnote}In case (i) of Fig.~\ref{fig:fasttav} we kept
$k_{12}=k_{11}$ in accordance with the experimental 
data \cite{english06} which indicate a difference between the values of $k_{2i}$ only.
\end{thebibliography}
\end{document}